\documentclass[preprint2]{aastex}

\shorttitle{Electron acceleration by a strong double layer}
\shortauthors{Dieckmann and Bret}

\begin{document}

\title{PIC simulation of a strong double layer in a nonrelativistic 
plasma flow: Electron acceleration to ultrarelativistic speeds}


\author{Mark E. Dieckmann\altaffilmark{1}}
\affil{Department of Science and Technology (ITN), Link\"oping University, 
SE-60174 Norrk\"oping, Sweden}

\and

\author{Antoine Bret}
\affil{ETSI Ind., Universidad de Castilla-La Mancha, 13071 Ciudad Real, Spain}
\email{Mark.E.Dieckmann@itn.liu.se}

\altaffiltext{1}{Visiting researcher at the Centre for Plasma Physics (CPP), 
Queen's University Belfast, BT7 1NN Belfast, U.K. }

\begin{abstract}
Two charge- and current neutral plasma beams are modelled with a 
one-dimensional PIC simulation. The beams are uniform and unbounded. 
The relative speed between both beams is 0.4c. One beam is composed 
of electrons and protons and one out of protons and negatively charged 
oxygen (dust). All species have the temperature 9 keV. A Buneman 
instability develops between the electrons of the first beam and the 
protons of the second beam. The wave traps the electrons, which form 
plasmons. The plasmons couple energy into the ion acoustic waves, which 
trap the protons of the second beam. A proton phase space hole grows, 
which develops through its interaction with the oxygen and the heated 
electrons into a rarefaction pulse. This pulse drives a strong ion 
acoustic double layer, which accelerates a beam of electrons to about 
50 MeV, which is comparable to the proton kinetic energy. The proton 
distribution eventually evolves into an electrostatic shock. Beams of 
charged particles moving at such speeds may occur in the foreshock of 
supernova remnant shocks. This double layer is thus potentially relevant 
for the electron acceleration (injection) into the diffusive shock 
acceleration by supernova remnants shocks.
\end{abstract}


\section{Introduction}

The blast shell of a supernova remnant (SNR) can accelerate charged 
particles to cosmic ray energies through diffusive shock acceleration 
by the SNR shocks \citep{Uchiyama,Reynolds,Hillas}. Particles are 
scattered upstream and downstream of a shock and some particles cross 
the shock repeatedly, gaining energy each time \citep{D1,D2,D3,Luke1}. 
Diffusive shock acceleration requires a seed population of particles 
with energies well in excess of the thermal ones, since only such 
particles can cross the shock repeatedly and be scattered efficiently 
by the magnetohydrodynamic waves on either side of the shock. The 
presence of shock-accelerated electrons is evidenced by the radio 
emissions of SNRs. It is not clear yet how the electrons can reach 
the energy threshold for diffusive shock acceleration; that requires that 
their kinetic energies are comparable to those of the ions. Some 
electrons must be accelerated locally, that is close to the SNR shock 
or close to one of its precursors \citep{Reynolds,Hillas}.

This injection of electrons is thought to be accomplished by the interplay 
of energetic ions with electrons and dust \citep{Luke2} by means of plasma 
instabilities \citep{Cargill}. The instabilities occur in the 
upstream plasma, just ahead of the SNR shock. This region is called the
foreshock. One source of the ion beams is the downstream plasma. Ions can 
leak through the shock boundary and outrun it \citep{Leaking}. The shock 
can also reflect a substantial fraction of the upstream ions. If the 
reflection is specular, then the ions can cross the foreshock at twice the 
shock speed. Reflected and leaked ion beams can be observed in situ at the 
Earth bow shock \citep{Foreshock}. Particle-in-cell (PIC) simulations of 
fast plasma shocks suggest, that they may also occur at SNR shocks when 
the upstream magnetic field is orthogonal to the shock normal 
\citep{Hoshino,Lem2,Chapman,Amano}. Precursor shocks move ahead of the 
main SNR shock. They can reach mildly relativistic speeds for the most 
extreme supernova explosions \citep{Kulkarni}. Such shocks can also reflect 
ions in the presence of a quasi-parallel magnetic field \citep{NewApJ}.

We focus in this work on ion beam instabilities in the foreshock, but
we do not model the shock itself. We rely on a foreshock model that 
has been widely used, e.g. in the works by \citet{H1,H2} and by 
\citet{TwoBeam,Die}. This model considers an ion beam, which moves 
through the upstream plasma, consisting of cool electrons and protons. 
The system is spatially uniform. Often a second, counter-propagating ion 
beam is introduced, which cancels the current of the first one. This 
second beam can be due to ions that have been reflected by a perpendicular
shock and which return to the shock, after they have been rotated by the 
upstream magnetic field \citep{McClements}. Here a beam of co-moving 
negatively charged oxygen ions cancels the current of a proton beam.

This foreshock model can be used if all processes develop on scales well 
below an ion Larmor radius and an inverse ion cyclotron frequency in the
case of a perpendicular shock. The foreshock model may be representative 
over larger scales otherwise, because ions can expand much farther if they
move along the magnetic field, as it is observed at the Earth's bow shock.
This foreshock model is not self-consistent, because it decouples the 
foreshock physics from that of the shock. It also assumes a plasma that 
is spatially uniform, while the shock reflected ion beams vary in their 
mean speed and temperature. However, we can examine with this model the 
acceleration of the upstream electrons as a function of the bulk parameters, 
such as the beam temperatures and mean speeds and the background magnetic 
field. We can also separate individual processes from the global shock 
dynamics, that involves coupling over many scales \citep{h3}. This 
simplifies their identification and interpretation. The periodic boundary 
conditions also allow us to reduce the box size, making PIC simulations 
computationally efficient even for realistic ion-to-electron mass ratios. 
The insight gained from such simulations can then guide us in the selection 
of plasma parameters for the more realistic and demanding PIC simulations 
of full shocks.

Simulation studies employing this foreshock model have revealed that 
proton beam-driven instabilites are not efficient electron accelerators.
An obstacle is the mass difference between the electrons and the protons. 
Most plasma instabilities saturate by accelerating the electrons. 
The fields are too weak to accelerate the ions and the heated electrons 
quench most of the known plasma instabilities, e.g. through Landau damping. 
Often only a few per cent of the ion energy is transfered to the electrons 
\citep{H1,H2,TwoBeam,Die}. Initially, this is also the case for ion beams 
that expand into the upstream plasma \citep{DM} prior to the formation of
a shock. Certain mechanisms \citep{VV,Kat} can, in principle, accelerate 
the electrons to relativistic speeds, but this remains yet to be demonstrated 
with self-consistent PIC simulations. However, the electrons, that have 
been accelerated by an instability like the Buneman instability (BI) 
\citep{Buneman}, are often spatially bunched, for example in form of phase 
space holes \citep{Roberts,Schamel1,Schamel2,TwoBeam}. Electron bunches or 
plasmons have a much higher inertia than electrons and they modulate the 
ions to give ion acoustic waves. The ion acoustic waves can trigger further 
processes, one of which we consider in this work. 
 
We investigate a plasma consisting of two spatially uniform beams. One 
beam is composed of protons and electrons and the second one of protons 
and oxygen carrying a single negative charge. Each of the beams is charge 
and current neutral. Both beams move at a relative speed of 0.4c. All 
species have the temperature 9 keV. These initial conditions are the 
simplest ones that take into account three particle species and they 
permit a first assessment of the impact of heavier ions. Section 2 
discusses in more detail these initial conditions and their potential 
relevance for a foreshock plasma. Section 3 describes the simulation 
results, which can be summarized as follows. A BI develops between the 
electrons and the protons of the second beam. The BI saturates by forming 
electron phase space holes and these plasmons feed wave energy into the 
ion acoustic wave \citep{Mendonca,Mendonca2}. Proton phase space holes 
develop, once the ion acoustic waves are strong enough to trap the beam 
protons. The electrons eventually thermalize and they reach a mildly 
relativistic thermal speed, which is stabilizing the system for some 
time. The BI plays no further role after the electron thermalization. We 
examine in detail a proton phase space hole that grows into being a proton 
density pulse with a negative density \citep{Infeld1,Infeld2}. The growth 
of this proton phase space hole may be caused by its interaction with the 
negative oxygen \citep{Bengt} or by the disruption of the electron current 
\citep{Smith}. The pulse triggers a strong ion acoustic double layer and 
it evolves subsequently into an electrostatic shock. The electron phase 
space structure and evolution of this double layer resembles that, found 
by a recent simulation study by \citet{Newman}. That study addressed the 
electron acceleration in the Earth's auroral ionosphere. However, our 
simulation yields the much stronger electron acceleration that is necessary 
for injecting the electrons into the diffusive shock acceleration at SNR 
shocks. The proton phase space distribution also differs from the standard 
picture of double layers outlined, for example, in the review papers by 
\citet{Smith} and by \citet{RaaduRas}. The trapped and free streaming 
protons in our simulation are well-separated in velocity, which is not 
generally the case for double layers. The substantial free energy stored 
in the difference of the mean velocities of both beams causes the extreme 
electron acceleration up to 50 MeV. We discuss our simulation results in 
section 4, we relate it to previous investigations of double layers and 
we discuss the necessary future research.

\section{The initial conditions and the simulation method}

The bulk plasma of a supernova blast shell can expand into the ambient 
medium at a speed of up to 0.1c-0.2c, at least during its early phase 
\citep{Vink,Fransson}. A fraction of the blast shell plasma may expand 
even faster. A plasma flow speed of 0.9c has, for example, been observed 
for a subshell ejected by a particularly violent supernova explosion 
\citep{Kulkarni}. Shocks form where the expanding plasma impacts on 
the ambient plasma, the interstellar medium. The typical shock speeds 
should be comparable to the few per cent of $c$ the main blast shell 
plasma has, but some precursor shocks may be considerably faster.
The detailed structure in particular of the fastest SNR shocks is 
unclear. We assume here, that the foreshock region is resembling at 
least qualitatively that of the well-known Earth bow shock 
\citep{Foreshock} and we focus on the ion foreshock. The ion foreshock
is characterized by the presence of ion beams that can propagate far
into the upstream.

However, we expect that 
a hot background of electrons is present in the foreshock, as PIC 
simulations of mildly relativistic shocks demonstrate \citep{NewApJ}.

We examine the interaction between two spatially uniform and unbounded
plasma beams, consisting of one negatively and one positively charged
species each. One beam consists of electrons and protons and the second
one of protons and oxygen carrying a single negative charge.

The simulation box frame is the reference frame, in which
both plasma beams move in opposite x-directions and with equal speed
moduli $v_b$. The simulation models only the x-direction, which allows us
to resolve simultaneously the ion scales and the electron scales at the
correct mass ratio.

\begin{figure}
\plotone{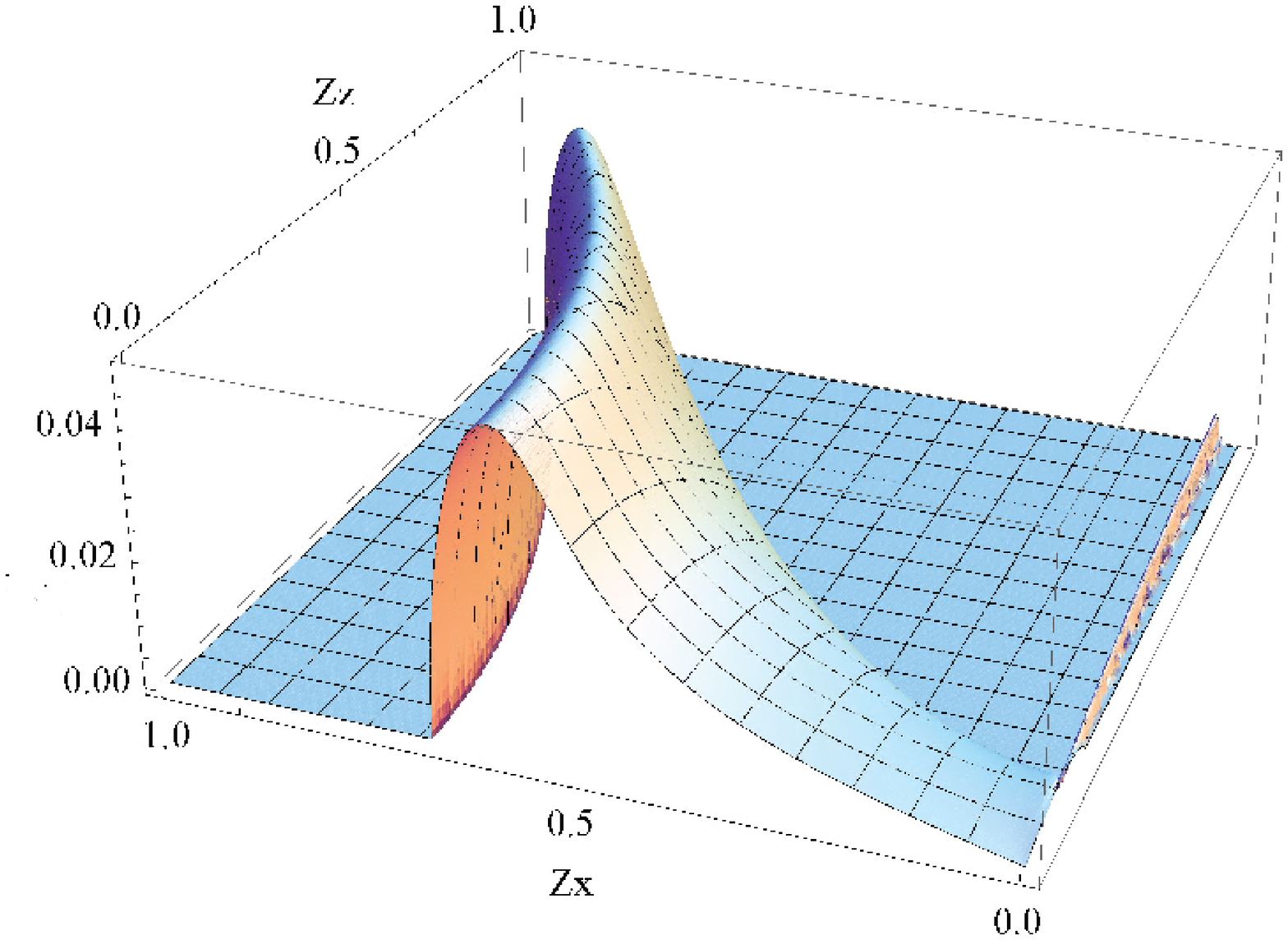}
\caption{The linear growth rate $\Omega_B / \Omega_e$ is given as the 
height and as a function of the flow-aligned wavenumber $Z_x = k_x v_b / 
\Omega_e$ and one perpendicular $Z_z = k_z v_b / \Omega_e$. The growth 
rate maximum $\Omega_B / \Omega_e = 0.049$ is found at $Z_z = 0$ and $Z_x 
= 0.59$, which is close to the $\Omega_{BC}/\Omega_e = 0.056$ and $k_u v_b 
/ \Omega_e = 0.5$ of the cold BI.}\label{fig1}
\end{figure}

The beam 1 consists of electrons with the plasma frequency $\Omega_e =
{(e^2 n_e/m_e \epsilon_0)}^{1/2}$, where $e$ and $n_e$ are the elementary
charge and the electron number density and $\epsilon_0$ is the dielectric
constant. The equally dense protons have the plasma frequency $\Omega_p =
{R_p}^{1/2} \, \Omega_e$, where $R_p = m_e / m_p$. Both species move at
the mean speed $<v> = -v_b$ with $v_b=c/5$. The beam 2 is composed of
protons with $\Omega_p$ and oxygen, which carries a single negative charge.
The oxygen plasma frequency is $\Omega_o = \Omega_p / 4$. The mean speed
of both species of beam 2 is $<v>=v_b$. All species have the temperature
$T=9.1$ keV, which gives the electron thermal speed $v_e = {(k_B T /
m_e)}^{0.5} = v_b / 1.5$. By neglecting the $T\neq 0$ and the heavy species
of each beam, the dispersion relation for the beam-aligned electrostatic
modes can be cast into the form
\begin{equation}
1 - \frac{\Omega_e^2}{{(\omega + k \, v_b)}^2} -
\frac{\Omega_p^2}{{(\omega - k \, v_b)}^2} = 0.
\end{equation}
The wavenumber and frequency in the reference frame of beam 1 of 
the most unstable wave are $k_u = \Omega_{e} / 2 v_b$ and $\omega_u = 
\Omega_{e}$. The Doppler-shift will reduce the frequency of the most
unstable wave to $\Omega_e/2$. The wavelength $\lambda_u = 2\pi / k_u$. 
The phase speed $v_{ph} = \omega_u / k_u \approx 2 v_b$ in the
reference frame of beam 1 or $v_b$ in the box frame. Replacing the oxygen
with electrons gives symmetric beams, which lets the phase speed of the
waves vanish in the box frame. An exact resonance between the plasmons and
the proton beam would not be possible. The estimates for $\omega_u$ and
$k_u$, as well as the growth rate $\Omega_{BC} / \Omega_e \approx ({3}^{1/2}
/ 2^{4/3})(\Omega_p / \Omega_e)^{2/3} \approx 0.056$ of the cold BI
\citep{Buneman} are fairly accurate. Figure~\ref{fig1} shows the numerically
obtained growth rate $\Omega_B / \Omega_e$ for our plasma parameters. This
growth rate map has been obtained by using a warm fluid model \citep{BretPRL}.
The growth rate map is symmetric around $v_b$, so we do not need to consider
$k_y$. The growth rate in the warm plasma model is lower than in the cold
one by about 10\%. The corresponding wave number is increased by 20\%. The
growth rate decreases for increasing $k_z$.

The PIC simulation method \citet{Dawson} solves the Vlasov-Maxwell
set of equations with the method of characteristics. It approximates
the phase space density $f(\mathbf{x},\mathbf{v},t)$ by an ensemble
of computational particles (CPs). The CPs correspond to phase space
volume elements and are not physical particles. The charge $q_{cp}$ 
and mass $m_{cp}$ of a CP can thus differ from those of the particles
it represents. The charge-to-mass ratio $q_{cp}/m_{cp}$ must, however,
equal that of the physical particles and the ensemble properties of
the CPs approximate well those of the physical plasma. The trajectories
of the CPs are prescribed by the Lorentz equation of motion. The
Maxwell-Lorentz set of equations can be normalized, which provides
us with results that are independent of the plasma density.

Let $e$ and $m_e$ be the elementary charge and the electron mass.

Our code is based on the numerical scheme outlined by
\citet{Eastwood} and it solves the Maxwell's equations for all components
of the electromagnetic fields together with the Lorentz equation, which
updates the three-dimensional vector of the relativistic momentum for an
ensemble of computational particles (CPs).

The relevance of double layers for solar and astrophysical particle 
acceleration has been proposed first by \citet{Alfven}. \citet{Lembege} 
have found double layers with PIC simulations of 
oblique shocks, which dissipated the electron energy. In our PIC
simulation the double layer emerges out of the BI and prior to the 
formation of the shock. A relativistic and monoenergetic electron beam 
is ejected by the double layer. The time evolution of the proton pulse 
implies, that the mean energy of this beam rises in time to about 50 MeV, 
which is a much stronger acceleration than that reported, for example, for 
previous PIC simulations of double layers \citep{Sato}. 

The nonrelativistic beam speed and the high temperature ensure that the 
BI grows fastest. The approximation of this system by a 1D PIC simulation, 
which enforces a monodirectional (plane) wave spectrum, may thus initially 
be justified. This geometry and the idealized initial conditions facilitate 
the interpretation of the physical processes and they permit us to resolve 
the relevant spatio-temporal scales. However, the growth rate of the
oblique modes is not negligible and the wave power is isotropized in time 
through weak turbulence \citep{Ziebell}. The PIC simulation may also become 
unrealistic when the electrons have been accelerated to relativistic speeds. 
This is, because of the unresolved higher-dimensional instabilities, the 
generation of energetic photons through synchrotron- and bremsstrahlung and 
the back-reaction of these processes onto the plasma 
\citep{Weibel,BretRel,Schlickeiser,Brems1,Brems2}. 

Charged dust and heavy ions exist in the supernova ejecta and in the 
interstellar medium \citep{Luke2,Meikle}. Their ionization state and 
their relative densities are unclear. It is typically only the dust, 
which has a negative time-dependent charge \citep{Kourakis,Verheest}, 
while the oxygen is positively charged \citep{Oxygen}. Our initial 
conditions are thus not necessarily representative for SNR blast shell 
plasmas. However, the purpose of the oxygen is to introduce a slower 
time scale \citep{Bengt} and also to obtain a resonance between plasmons 
and the ion acoustic wave, which facilitates their coupling~\citep{Mendonca}. 
Both conditions can probably also be achieved by heavier ions, by a mix of 
ions or by dust.

The beam speed of 0.4c is too high for typical SNR bulk flows, which can
reach about 10-20\% of the
speed of light $c$ \citep{Vink,Fransson}. However, the shock-reflected
ion beams can outrun the shock. Some components of the expanding plasma
may also move faster than the mean speed of the SNR ejecta. Precursor
shocks of particularly strong SNRs may reach speeds of up to 0.9c
\citep{Kulkarni}. We select this beam speed, because it facilitates the
processes we consider and because it allows us to estimate the maximum
energy, which the electrons could gain in typical SNR flows through the
plasma instabilities discussed here. The beam speed is also low enough
to favor the growth of the BI rather than that of the multi-dimensional
filamentation instability \citep{Lee}.

The simulation box length is 
$L \approx 1170 \lambda_u$. It is resolved by $N_g = 7 \cdot 10^4$ grid 
cells, each with the length $\Delta_x \approx 0.31 \lambda_D$. The electron 
Debye length is $\lambda_D = v_e / \Omega_e$. Each of the four species is 
represented by 250 CPs per grid cell, giving a total of $N_p = 1.75 \cdot 
10^7$ per species. The system is followed in time for $t_{S}\Omega_e 
\approx 2600$, which is subdivided into $10^5$ time steps $\Delta_t$. The 
beams propagate the distance $t_S v_b < L/5$ and even a light pulse can 
cross $L$ just once. Effects due to the periodic boundary conditions of 
the simulation box can not occur. 

\section{Simulation results}

\begin{figure}
\plotone{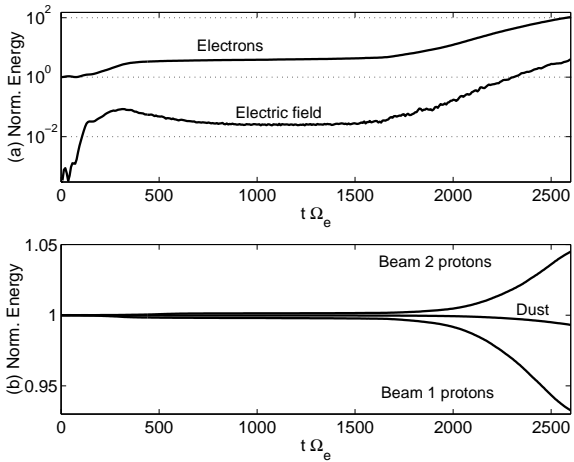}
\caption{The electrostatic energy density $E_F$ and the kinetic energy 
densities $E_i$ of the four particle species $i$: Panel a) shows $E_F$
and the $E_1$ of the electrons of beam 1, both normalized to $E_1(t=0)$. 
Panel b) plots the $E_2$ of the protons of beam 1, the $E_3$ of the 
protons of beam 2 and the $E_4$ of the oxygen of beam 2, which are all
normalized to their respective initial value.}\label{fig2}
\end{figure}

The electrostatic BI develops between the electrons of beam 1 and the 
protons of beam 2. However, the secondary instabilities give rise to 
electric fields, which will let all particle species interact. 
Figure~\ref{fig2} displays the time evolution of the relevant energy 
densities, which are the energy density $E_F(t) = \epsilon_0 N_g^{-1}
\sum_{j=1}^{N_g} E_{x}^2(j\Delta_x,t)$ of the electrostatic field and 
the $E_i(t) = N_g^{-1}\Delta_x^{-3} \sum_{j=1}^{N_p} m_i c^2[\Gamma (v_j) 
- 1]$, where the summation is over all particles $j$ of the species $i$ 
with the mass $m_i$ and a speed $v_j(t)$ giving $\Gamma (v_j) = 
{(1-v_j^2/c^2)}^{-1/2}$. 

The $E_F \propto E_x^2$ grows in the interval $t\Omega_e < 100$ at an 
exponential rate $2 \Omega_i \approx 0.06$, which is below $2 \Omega_B
=0.098$. The growth rate $\Omega_B$ is calculated for a sine wave. The 
$L \gg \lambda_u$ and the broad $k_x$ spectrum of unstable waves in 
fig.~\ref{fig1} implies, however, that we integrate over a mixture of 
waves, all of which have a growth rate $\le \Omega_B$ and the initial
growth of $E_F$ cannot be compared directly to $\Omega_B$.
The electron trapping lets $E_F(t)$ saturate at $t \Omega_e \approx 100$ 
and it causes the growth of $E_1(t)$. Thereafter $E_F(t)$ decreases and 
reaches a meta-equilibrium at $t\Omega_e \approx 10^3$. The $E_1(t)$ is 
almost constant until $t\Omega_e \approx 1500$, when it and $E_F(t)$ 
start to grow again. All ion beams have lost less than 1\% of their 
energy prior to $t\Omega_e \approx 1500$, confirming that the BI is not 
an efficient electron accelerator. Then $E_2(t)$ and $E_3(t)$ start to 
change and even $E_4(t)$ has decreased by 1\% at $t\Omega_e \approx 2600$.
 
\begin{figure}
\plottwo{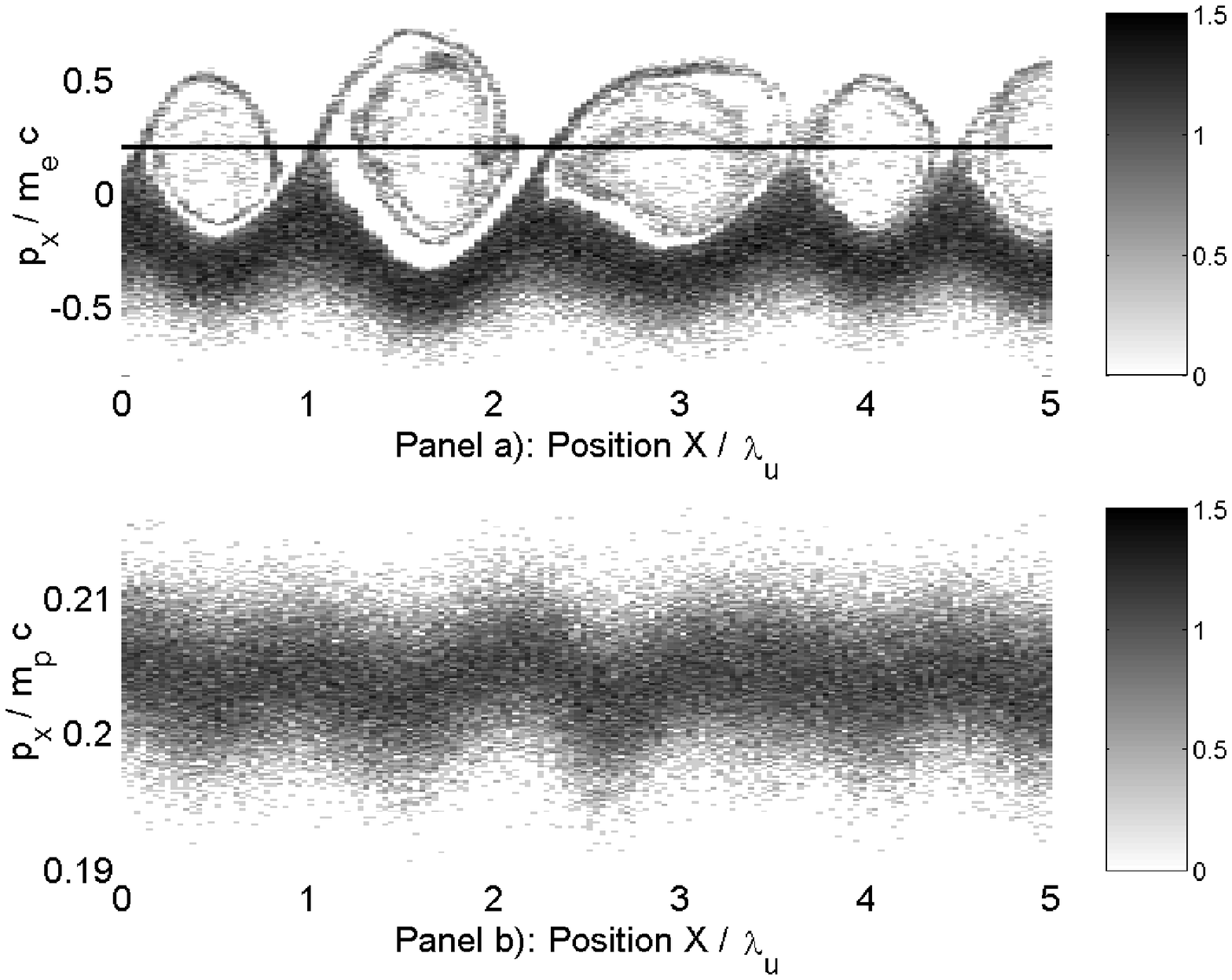}{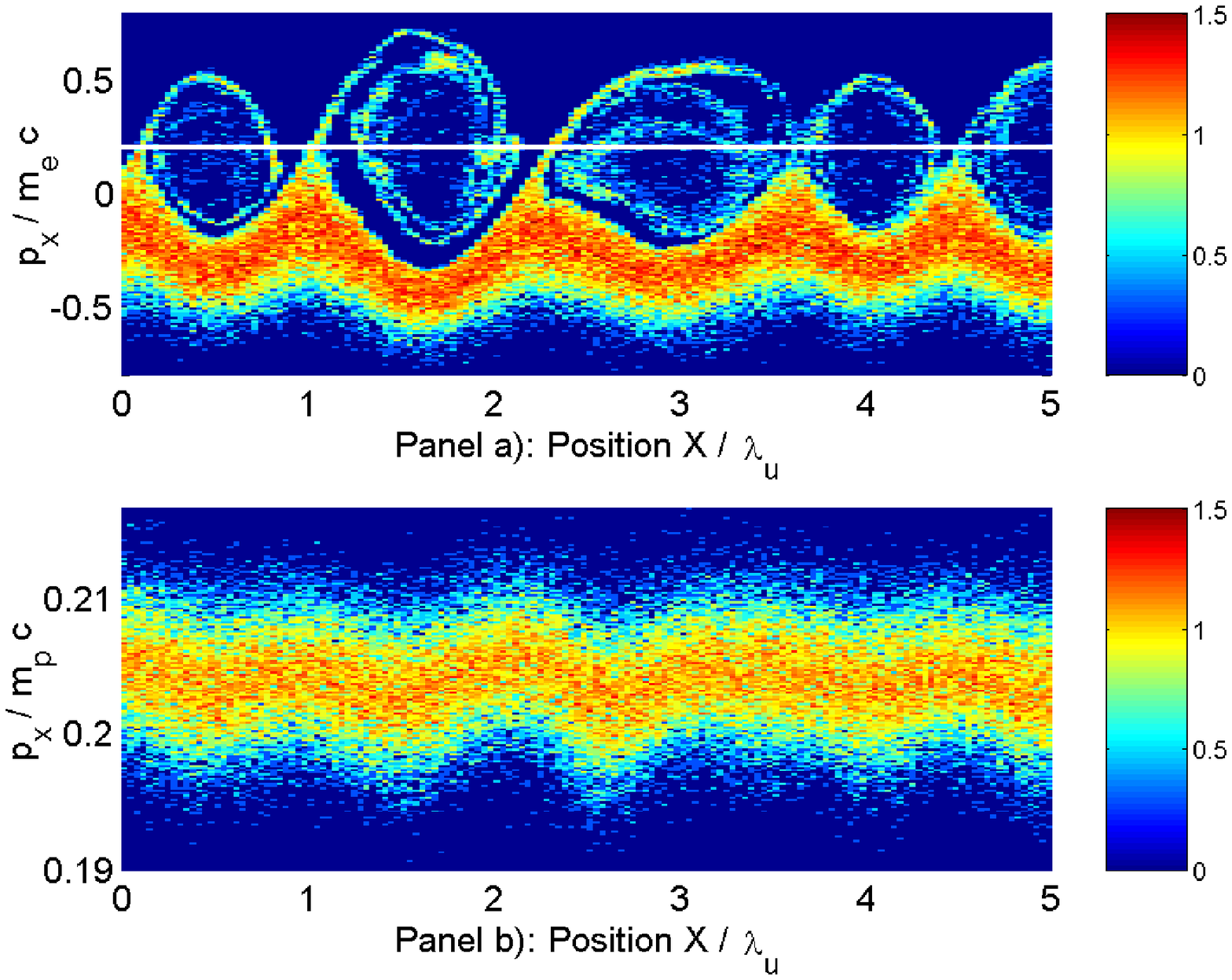}
\caption{The phase space distribution of the electrons (a) and of 
the protons of beam 2 (b) at $t\Omega_e = 145$: The colour scale
shows the 10-logarithm of the number of CPs. The electrons form phase
space holes, which are coalescing. The horizontal line corresponds
to the mean momentum $v_b/{(1-v_b^2/c^2)}^{1/2}$ of beam 2. The protons 
show velocity modulations that are spatially correlated with the electron 
phase space holes.}
\label{fig3}
\end{figure}

Figure~\ref{fig3} shows the phase space distributions of the trapped 
electrons and of the protons of beam 2 at $t\Omega_e = 145$ in a 
subinterval of the simulation box. The trapped electrons gyrate around a 
potential that is moving with the mean speed $v_b$ of beam 2, which is 
also the phase speed of the Buneman wave. Only a fraction of the electrons 
is trapped. The electrons of the untrapped bulk population move on 
oscillatory phase space paths. The electron phase space hole in the 
interval $0 < x / \lambda_u < 1$ has the expected size, while the electron 
phase space hole in the interval $2 < x / \lambda_u < 3.5$ is larger. 
Electron phase space holes increase their size by coalescence and their 
mutual interaction will deform them \citep{Roberts}. The mean momentum 
of the proton beam is weakly modulated. The initially periodic train 
of electron phase space holes will coalesce, until only a dissipative 
equilibrium and solitary electron phase space holes remain 
\citep{Schamel1,Schamel2}. 

\begin{figure}
\plottwo{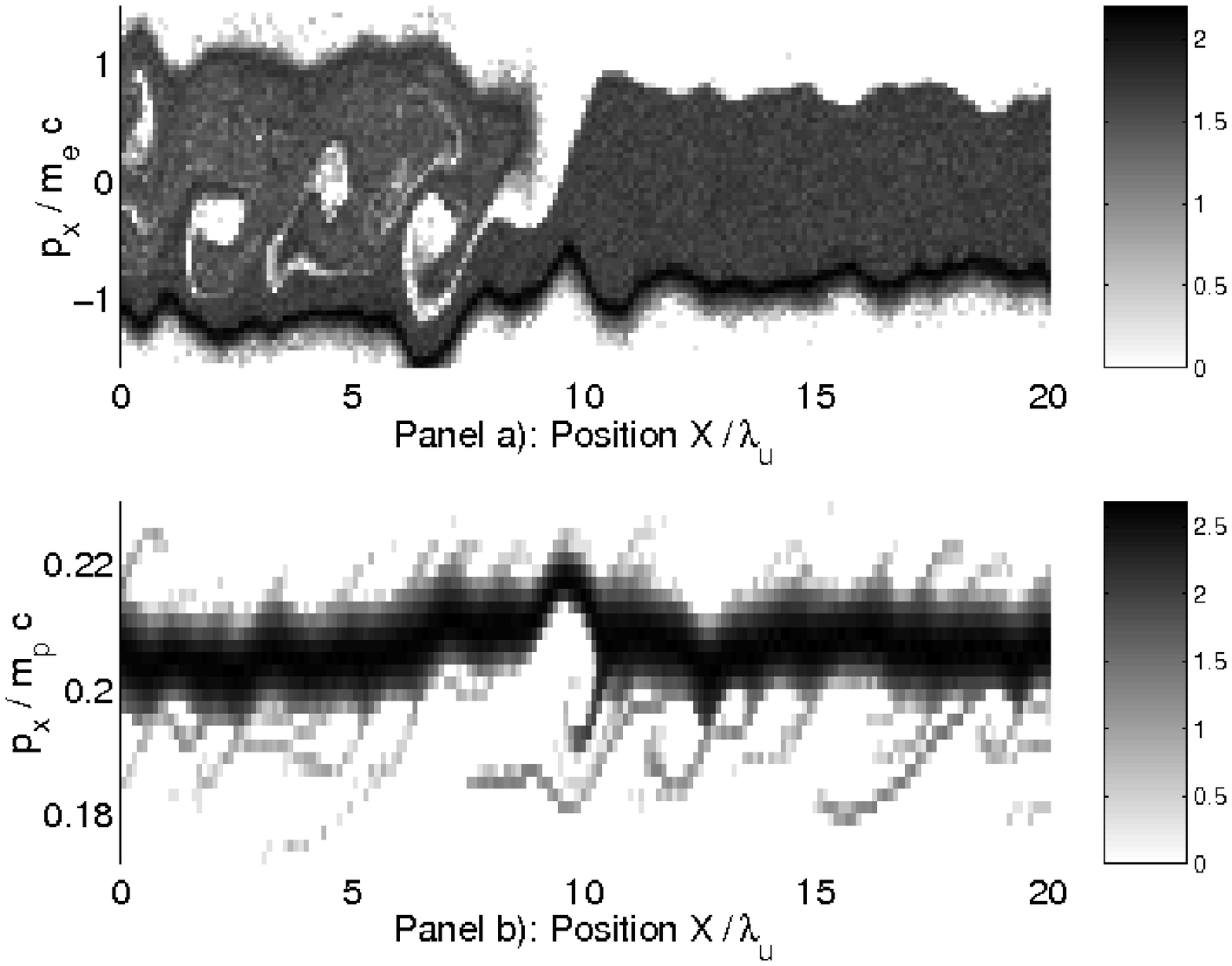}{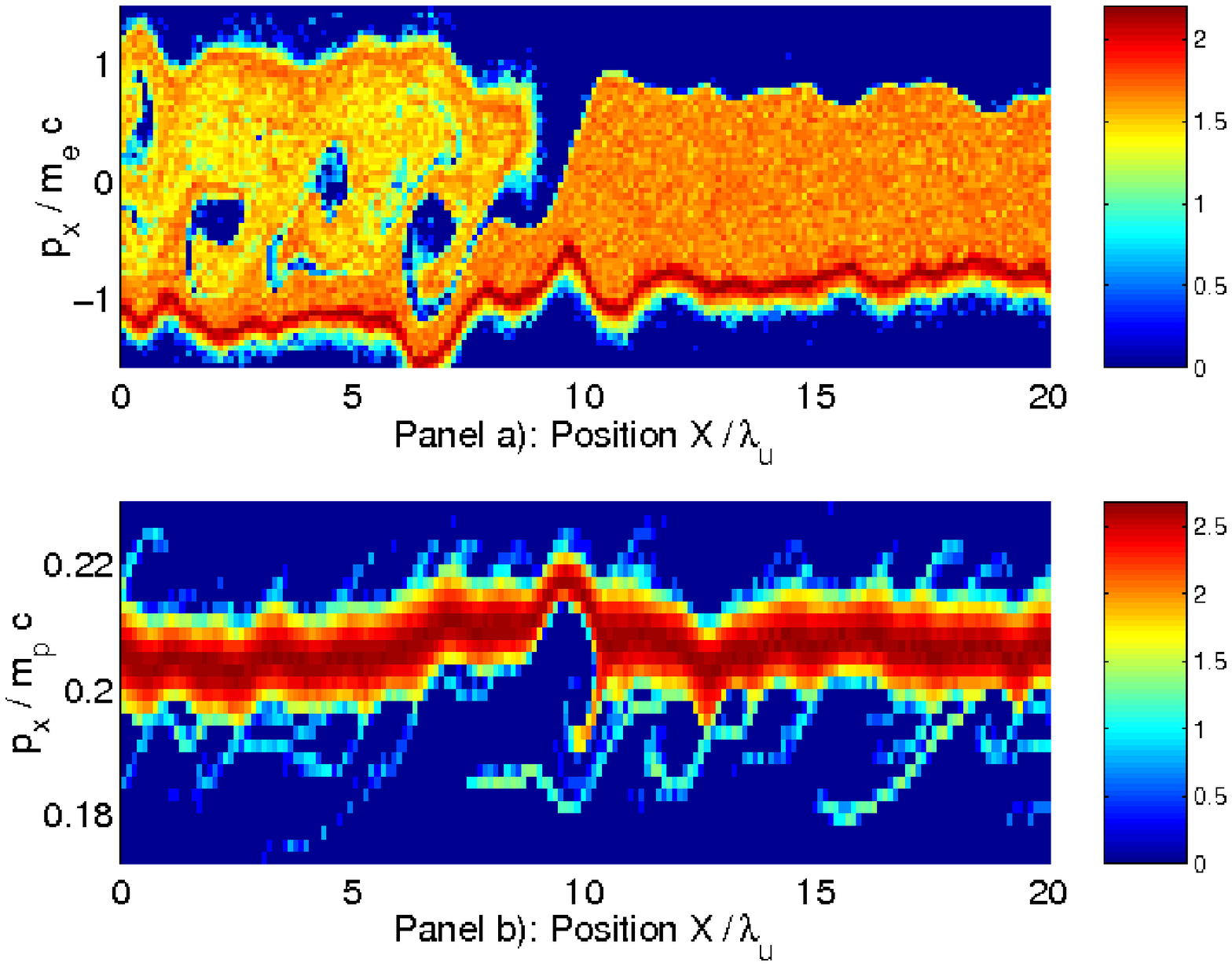}
\caption{The phase space distribution of the electrons (a) and of 
the protons of beam 2 (b) at $t\Omega_e = 10^3$: The colour scale
shows the 10-logarithm of the number of CPs. The electrons display
a plateau distribution for $x/\lambda_u > 10$ and vortices for 
$x/\lambda_u < 8$. The plateau is split at $x/\lambda_u \approx 9$.
The proton beam shows tenuous filaments, which are protons trapped 
in the ion acoustic waves. A phase space hole is forming at 
$x/\lambda_u \approx 9$.}\label{fig4}
\end{figure}

Figure~\ref{fig4} confirms this. A simulation box interval is selected at 
the time $t\Omega_e = 10^3$, in which an electron depletion at $x/\lambda_u 
\approx 9$ separates the turbulent region $x/\lambda_u < 8$ from a smooth 
plateau at larger $x$. The latter constitutes a dissipative equilibrium. 
The supplementary movie 1 follows the time-evolution of the electron
phase space distribution at the position of the electron depletion. The
color scale is the 10-logarithm of the number of CPs. Note that the absolute 
position in the box is given in movie 1, while fig.~\ref{fig4} uses a
relative position. The movie 1 demonstrates that it is this density 
depletion, which is responsible for the turbulent phase space region. A 
spatial correlation is visible in the fig.~\ref{fig4} between the electron 
depletion and a phase space hole in the protons of beam 2. This explains 
why the electron depletion is stationary in the reference frame of movie 1, 
which moves with $v_b$ in the box frame. The electron temperature is now 
relativistically high, by which the linear damping of the ion acoustic 
waves is reduced and these modes can accumulate wave energy. The thermal 
anisotropy of the electrons causes the growth of the Weibel instability 
\citep{Weibel}, but it is suppressed here by the 1D geometry. 

\begin{figure}
\plotone{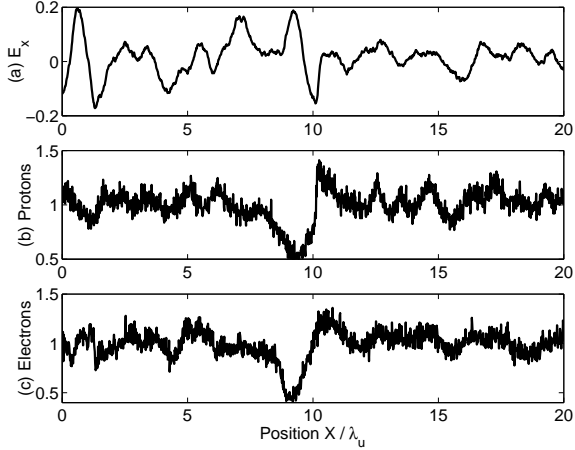}
\caption{The normalized electrostatic field $eE_x/(\Omega_e c m_e)$ (a), 
the proton density of beam 2 (b) and the electron density (c). The 
electric field shows a bipolar oscillation at $x/\lambda_u \approx 9$, 
which modulates the densities of the protons and electrons. The time is 
$t\Omega_e = 10^3$.}\label{fig5}
\end{figure}

Figure~\ref{fig5} plots the $E_x$, the density of the proton beam 2 and 
that of the electrons at the same location and time as fig.~\ref{fig4}. 
The electric field at $x/\lambda_u \approx 9$ is attractive for the 
protons and repels the electrons, which is in agreement with the observed 
phase space structure in fig. \ref{fig4}. The electric field amplitude at 
this position is not unusually large, while the plasma density depletion 
is. The oxygen density is unaffected at this time, as we show below. 

\begin{figure}
\plottwo{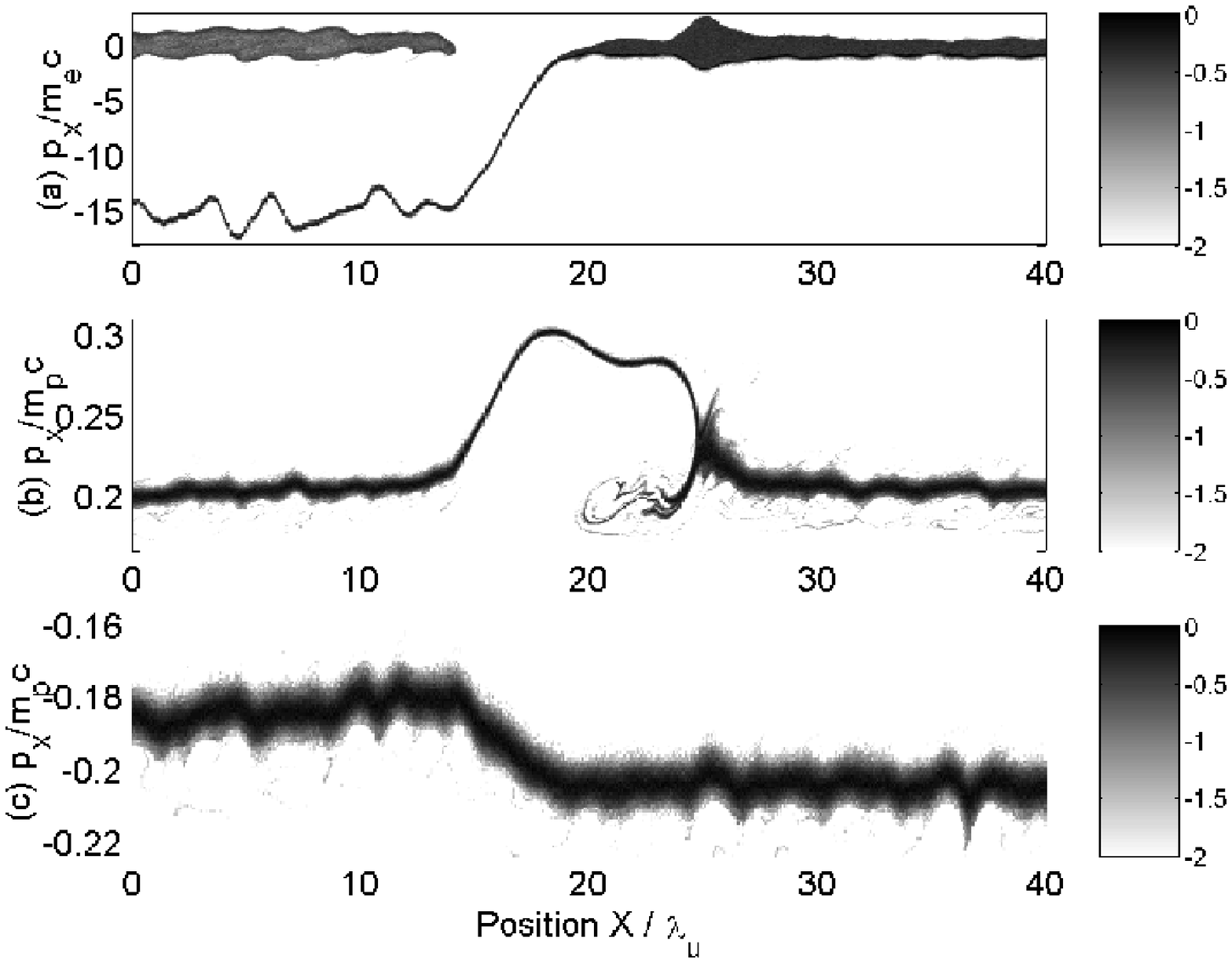}{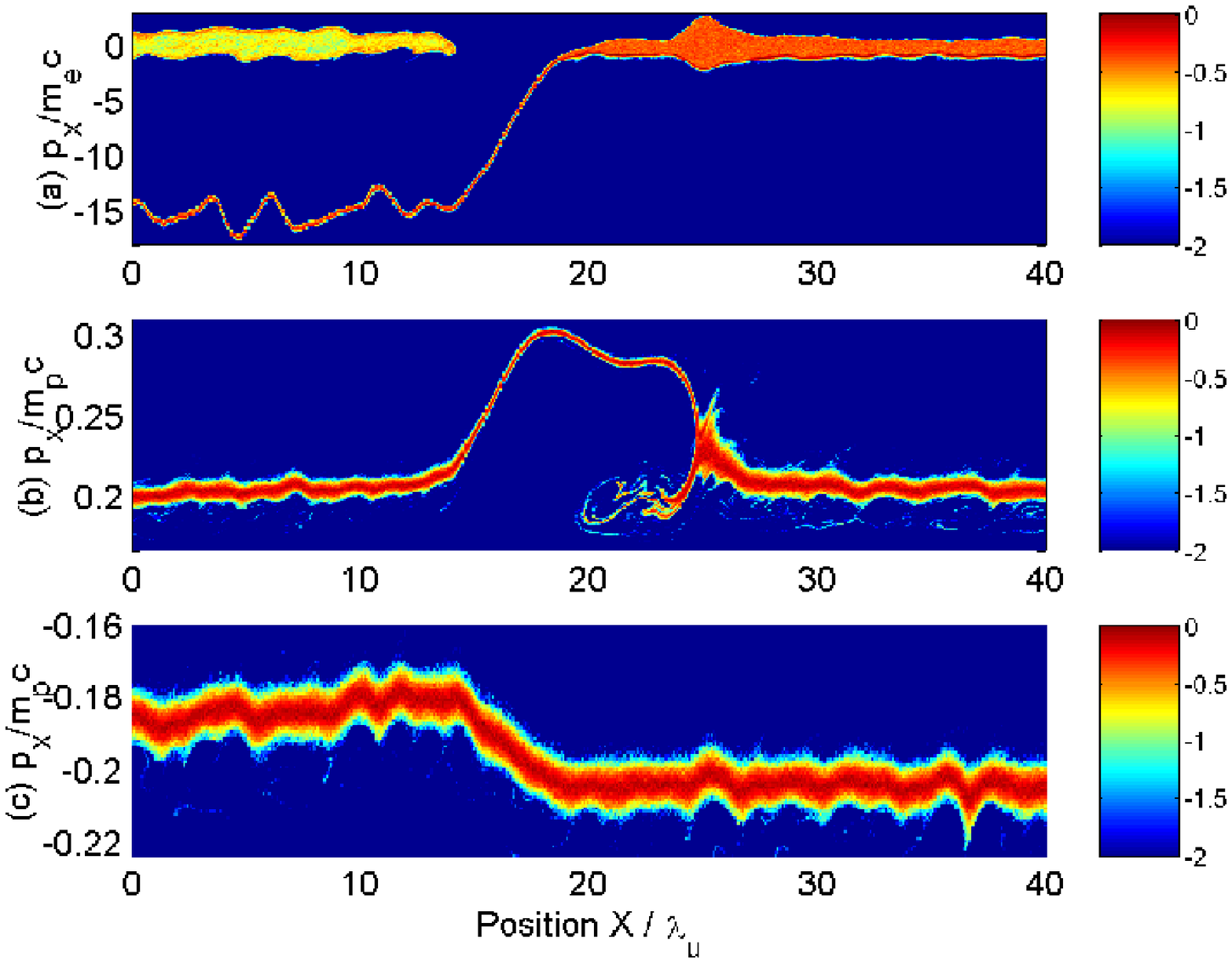}
\caption{The phase space distributions at $t\Omega_e = 2000$ of the
electrons (a), the protons of beam 2 (b) and of beam 1 (c). The colour
scale is the 10-logarithm of the number of CPs. The electron distribution 
evidences a double layer in the interval $15 < x / \lambda_u < 20$. The
Double layer is maintained by the proton pulse moving with the beam 2,
which is strong enough to modulate the protons of beam 1.}
\label{fig6}
\end{figure}

The component energies in fig.~\ref{fig2} demonstrate that the equilibrium
between $E_F(t)$ and $E_1(t)$ breaks at $t\Omega_e \approx 1500$ and that
a secondary instability sets in after that time. An examination of the 
full phase space data shows, that this secondary instability is spatially 
localized. Two similar, but spatially well-separated structures develop 
in the phase space distribution of the protons of beam 2 in the box. That 
two structures rather than one grow suggests, that they develop with a 
reasonable probability and that their growth is not accidental. 
Figure~\ref{fig6} displays the larger of these two structures at $t 
\Omega_e = 2000$. A different sub-interval of the box than in the 
fig~\ref{fig4} is investigated, since the proton phase space structures 
convect with $v_b$. We refer to this proton structure as a pulse. This 
pulse constitutes a density depletion \citep{Infeld1,Infeld2}. The 
supplementary movie 2 animates in time the development of the pulse out 
of the proton phase space hole. The color scale shows the 10-logarithm of 
the number of CPs. The movie 2 displays absolute box coordinates and its 
reference frame moves with $v_b$ through the box. It demonstrates that the 
proton phase space hole is stationary for a long time, before it develops 
into a rapidly growing pulse. It is this pulse that is responsible for the 
energy changes in the fig~\ref{fig2}.

The electron phase space distribution reveals in fig.~\ref{fig6} a double 
layer that is driven by the pulse. This double layer accelerates on $5 
\lambda_u$ a beam of electrons out of the bulk to about 7.5 MeV. The 
movie 1 demonstrates that the mean energy of the electron beam grows in 
time, while the beam temperature remains unchanged. The electron beam 
then interacts with the bulk electrons through a two-stream instability, 
which is responsible for the large beam momentum oscillations. The electric 
field of the double layer is strong enough 
to decrease the mean speed modulus of the proton beam 1, which supplies 
the energy for the electron acceleration. The movie 2 is stopped, when 
the protons of beam 1 and 2 start to mix by the formation of an 
electrostatic shock. The movie 1 is stopped, when the electron beam energy 
exceeds the one resolved by the movie window.

\begin{figure}
\plotone{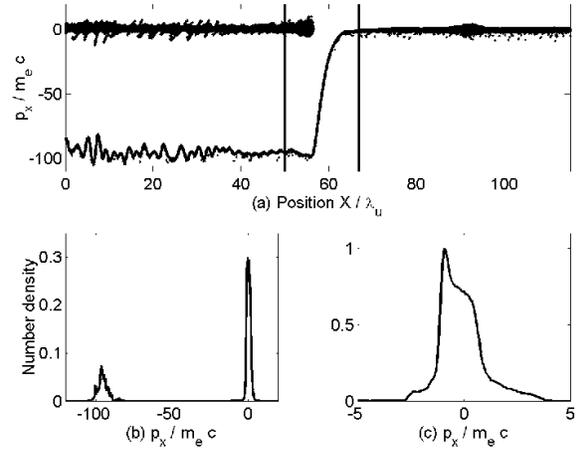}
\caption{The electron distribution at $t\Omega_e = 2600$. The phase
space distribution in (a) demonstrates that the width of the double
layer is still $5\lambda_u$, but that the electrons reach now 50 MeV.
The electron phase space distribution is intergrated over the left 
interval $x/\lambda_u < 50$ in (a) and the number density is plotted 
in (b). Two beams are visible. The integral over $x/\lambda_u > 66$
in (c) shows a single electron beam.}
\label{fig7}
\end{figure}

The electron distribution at $t\Omega_e = 2600$ is illustrated by
fig~\ref{fig7}. The double layer has not spatially expanded but the 
electrons are now accelerated to 50 MeV in the box frame, exceeding even 
the 20 MeV energy of the protons in the same reference frame. The total 
density of the ultrarelativistic beam in fig.~\ref{fig7}b) is about 40\% 
of the density of the bulk electrons in fig.~\ref{fig7}c). The momentum 
distribution of the bulk electrons is practically identical to that in 
the fig.\ref{fig4}. This ultrarelativistic electron beam would drive the 
unresolved and faster-growing oblique mode instability \citep{BretRel}, 
rather than the two-stream instability evidenced by the movie 1. The 
type of instability should, however, not be important for the evolution
of the electron double layer because it forms well behind it.

\begin{figure}
\plotone{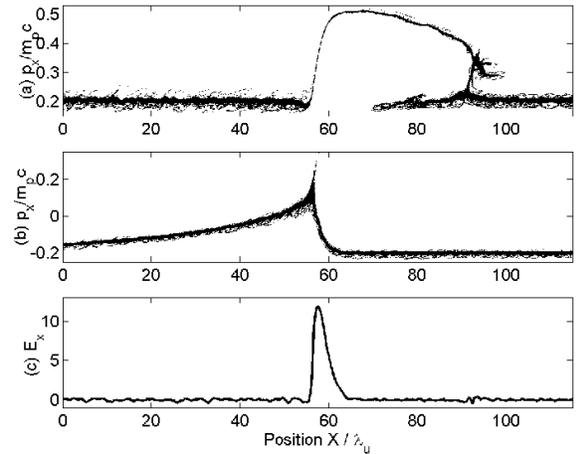}
\caption{The proton distribution at $t\Omega_e = 2600$ of beam 2 (a) 
and of beam 1 (b). Panel (c) shows the normalized electric field 
$eE_x/(\Omega_e c m_e)$. A large electrostatic shock is forming at 
$x / \lambda_u \approx 55$, which involves the protons of both beams. 
An electrostatic shock involving only the protons of beam 2 is driven 
by the rapid expansion of the pulse at $x / \lambda_u \approx 92$.}
\label{fig8}
\end{figure}

Figure~\ref{fig8} displays the phase space distributions of both proton
beams in the same spatial interval as fig.~\ref{fig7}. The proton 
distributions of both beams have merged and a large electrostatic 
shock forms at $x/\lambda_u \approx 55$. A second electrostatic shock 
involving only the protons of beam 2 occurs at $x/\lambda_u \approx 92$. 
The movie 2 illustrates that the second shock is driven by the rapid
expansion of the proton density pulse. It is remarkable that some protons 
of beam 2 have increased their $p_x$ momentum by a factor 2.6 through the 
electrostatic field, which is also shown by fig.\ref{fig8}. The electron 
double layer is thus also a proton accelerator. The electrostatic field 
of the double layer does not change its sign, which contrasts the bipolar 
fields of phase space holes, like in fig~\ref{fig5}.

\begin{figure}
\plotone{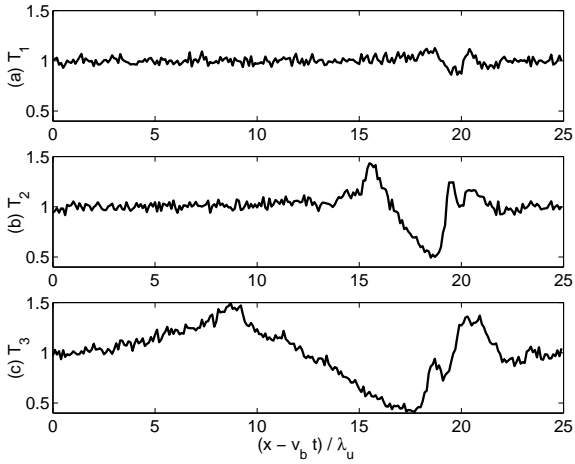}
\caption{The oxygen density at the times $t\Omega_e = 1000$ (a), at 
$t\Omega_e = 1600$ (b) and $t \Omega_e = 2000$ (c) in the reference 
frame convected with the proton beam 2. The location of the original 
proton phase space hole is at $(x-v_b t)/ \lambda_u \approx 20$. The 
density is integrated over 3 grid cells to reduce the noise. The density 
in (a) shows a density depletion of the order of the number density 
fluctuations. The density in (b) is modulated by about 50\% and it has 
expanded in size in (c). The density varies approximately linearly over 
a limited interval in (b) and in (c).}\label{fig9}
\end{figure}

Solitary phase space holes are typically time-stationary in one dimension
\citep{Roberts,Schamel1,Schamel2} and the question arises, why the proton 
phase space holes grow here into such powerful rarefaction pulses. The 
proton phase space hole is stable over a long time and it then suddenly 
turns into a pulse, which suggests an effect that develops on a slower 
time scale. The electric field of the proton phase space hole will gradually 
displace the oxygen ions. Figure \ref{fig9} displays the density distribution 
of the oxygen ions at the three times $t \Omega_e = 1000$, $t \Omega_e = 
1600$ and $t \Omega_e = 2000$. The frame is moved with the beam 2 and the 
proton phase space hole is located at $(x-v_b t) \approx 20$ in the chosen 
frame. The oxygen depletion at $t\Omega_e = 10^3$ is comparable to the 
statistical fluctuations of the number of CPs. The integration of the 
density over 3 grid cells reduces these statistical fluctuations by the 
factor $\sqrt{3}$, but it keeps the density depletion value unchanged. The 
proton phase space hole grows into the pulse at $t\Omega_p \approx 1500$. 
Figure \ref{fig9}b) reveals that the oxygen density modulation is significant 
at around this time. It is thus likely that the sudden growth of the proton 
phase space hole is a consequence of this oxygen depletion. As the proton 
pulse grows, the structure and amplitude of the oxygen depletion remain 
qualitatively unchanged, but the depletion expands in space. This may reflect 
the observation from the movie 2 that the shape of the proton pulse is 
practically unchanged until $t\Omega_e \approx 2000$. The proton pulse
and the oxygen depletion are intertwined.

The proton beam 1 and the electrons in our simulation have a significant 
mean speed in the reference frame of the proton phase space hole and they
could provide the energy for the growth of the proton pulse. The movie 1 
reveals that the proton phase space hole gives rise to vortices in the 
electron phase space distribution and the electron flow is turbulent, 
which excerts a drag on the proton phase space hole. A coupling between 
the proton phase space hole and the streaming electrons is established,
which can also trigger or accelerate the growth of the phase space hole 
\citep{Dupree}. The simulation further evidences that the proton pulse 
grows at the expense of the kinetic energy of the proton beam 1. 

\section{Discussion}

This work has examined secondary instabilities triggered by the Buneman
instability (BI) \citep{Buneman} with a PIC code \citep{Eastwood}. One 
spatial x-direction has been resolved, which implies that the wave 
spectrum driven by the BI is monodirectional. This limits the realism
of the simulation but it firstly allows us to resolve all relevant 
spatio-temporal scales and, secondly, it facilitates the identification
of the relevant physical processes. The large box has ensured that the 
periodic boundary conditions have not affected the results. We have 
considered two unmagnetized, interpenetrating, uniform and unbounded 
beams that have been composed of three particle species. The beam 1 has 
contained protons and electrons and the beam 2 has been formed by protons 
and by oxygen with a single negative charge. Each beam has been charge-
and current neutral. The relative speed between both beams has been 
set to 0.4c and the initial temperature of all species has been selected 
to be 9.1 keV.
 
The BI has saturated by the trapping of electrons \citep{Roberts}, which 
groups the electrons into phase space holes that give rise to charge 
density modulations. The latter can behave as quasi-particles and 
are thus referred to as plasmons. We could match the mean (group) speed 
of these plasmons with that of the protons of beam 2 by selecting negatively 
charged oxygen rather than electrons. Such a resonance facilitates the 
coupling of the plasmon energy into the ion acoustic wave mode 
\citep{Mendonca}. 

Initially the equilibrium between the electric fields and the electrons 
has been preserved \citep{Schamel1,Schamel2} and the ion beams had lost 
only a small fraction of their energy; the electrons moved consequently 
only at mildly relativistic speeds. The thermal anisotropy of the electron
distribution would trigger an instability similar to that proposed by 
\citet{Weibel}, resulting in the growth of magnetic fields and 
multi-dimensional field structures that cannot be represented by a 1D
PIC simulation. The Weibel instability may, however, not be important. 
The proton phase space holes form simultaneously with the collapse of 
the electron phase space holes that gives rise to the thermal anisotropy. 
The electromagnetic fields of the Weibel instability may not be strong 
enough to affect the interplay between the already present proton phase
space holes and the oxygen and might be negligible. 

As the time progressed, two large density rarefaction pulses 
\citep{Infeld1,Infeld2} have developed in the protons of the beam 2. We 
have examined the larger one in more detail. The long time between the 
formation of the proton phase space hole and the pulse has suggested that 
it has been caused by the oxygen response. We have found that the proton 
density pulse grows, once the electric field of the proton phase space 
hole has displaced a substantial fraction of the oxygen ions. A related 
effect has been observed by \citet{Bengt}. There it was shown that an 
electron phase space hole was ejected by a density depletion in the 
background ions, which its electric fields excavated. Here, the consequence 
of the oxygen depletion is not the acceleration of the phase space hole but 
its growth. The electrons may also contribute to the growth of the proton 
phase space hole \citep{Dupree}.

The growth of the proton density pulse has resulted in an electron double 
layer that accelerated the electrons to ultrarelativistic speeds. To the 
best of our knowledge this is the first time that such an electron 
acceleration could be observed in PIC simulations of nonrelativistic beam 
instabilities. The electrons have reached a peak energy of 50 MeV. This 
immense electron acceleration implies, that the 1D PIC simulation geometry 
is not appropriate any more. Firstly, the electrons moving at such speeds 
drive the oblique mode instability \citep{BretRel}, rather than the 
two-stream instability. The oblique modes are excluded by the simulation 
geometry. Secondly, the extreme electron temperature anisotropy resulting 
from the thermalization of the ultrarelativistic electron beam would drive 
a very strong Weibel-type instability. The subsequent magnetic field growth 
would imply that electrons emit synchrotron radiation on top of the 
bremsstrahlung \citep{Brems1,Brems2}. The balance of these processes 
\citep{Schlickeiser} requires a three-dimensional geometry and a method to 
deal with the radiation, which is not possible with our PIC code. 

At the same time, this electron double layer constitutes a plasma structure
that can easily accelerate the electrons to a speed, which allows them to
enter the diffusive shock acceleration mechanism. This could be achieved with
beam speeds that are not much higher than the flow speeds of supernova
blast shells \citep{Vink,Fransson}. The oxygen ions we have modelled can 
also be found in such plasmas, although not in the high density and in the 
ionisation number we have assumed here \citep{Oxygen}. Our negative charge 
is more representative of dust, which we also find in supernova blast shells
\citep{Meikle}. Our findings can thus not be applied directly to a supernova 
scenario. However, the formation of the electron double layer in the 
presence of heavier ions clearly demonstrates the need to introduce these 
into PIC simulations, which so far only consider one ion species 
\citep{Amano,Chapman}. 

Future simulation work must address the minimum speed, at which the double 
layer can form. This has to be done in form of parametric simulation 
studies, because it is not clear for which combinations of the beam 
speeds and temperatures the double layers can form. Existing analytic 
studies \citep{Verheest} do not cover the case we consider here. The
effects of the obliquely propagating modes driven by the initial beam
instability have to be examined, as well as those due to the Weibel
instability driven by the thermal anisotropy of the electron distribution
after the initial BI has saturated. This requires at least two-dimensional 
PIC simulations. Finally, other ion species and plasma compositions have 
to be examined to assess how easily double layers can form. This is 
important, because the ion composition is not easily obtained \citep{Luke2} 
and not unique.

The modelling of double layers and their consequences is also of interest
for other astrophysical scenarios. Ion double layers, which are characterized 
by ion distributions that resemble the electron distribution here, may 
regulate the mass inflow onto neutron stars \citep{Layer1} and accelerate 
particles in accretion discs \citep{Layer2}. They may also contribute to 
solar coronal ejections \citep{Alfven}. 

\acknowledgments
This work has been financially supported by the visiting scientist scheme
of the Queen's University Belfast, by the Swedish research council 
Vetenskapsr\aa det and by the projects FTN 2003-00721 of the Spanish 
Ministerio de Educaci\'{o}n y Ciencia and PAI-05-045 of the 
Consejer\'{i}a de Educaci\'{o}n y Ciencia de la Junta de Comunidades de 
Castilla-La Mancha. The Swedish High-Performance Computer Center North 
(HPC2N) has provided the computer time and support.

\end{document}